\documentclass[12pt]{article}
\usepackage{amssymb}
\begin{document}
\begin{center}
\textbf{ON THE CARDY-VERLINDE ENTROPY FORMULA IN VISCOUS COSMOLOGY}

\bigskip
\bigskip
\bigskip

I. Brevik\footnote{E-mail address:  iver.h.brevik@mtf.ntnu.no}

\bigskip
\bigskip
Division of Applied Mechanics,\\Norwegian University of Science and Technology,\\
N-7491 Trondheim, Norway\\

\bigskip
\bigskip
S. D. Odintsov\footnote{E-mail address:  odintsov@ifug5.ugto.mx. On leave
from Tomsk State Pedagogical University, Tomsk, Russia.}

\bigskip
\bigskip
Instituto de Fisica de la Universidad de Guanajuato, Lomas de Bosque\\
103, Apdo. Postal E-143, Leon, Gto., Mexico\\

\bigskip
\bigskip

Revised version,  December 2001
\end{center}

\bigskip

\begin{abstract}

The results of the paper of Verlinde [hep-th/0008140], discussing the 
holographic 
principle in a radiation dominated universe, are extended when allowing 
the cosmic 
fluid to possess a bulk viscosity. This corresponds to a non-conformally 
invariant theory. The generalization of the 
Cardy-Verlinde entropy formula to the case of a viscous universe seems 
from a formal point of view to be possible, although we question on physical grounds some elements in this kind of theory,
 especially the manner in which the Casimir energy is evaluated. 
Our discussion suggests that for non-conformally invariant theories the
holographic definition of Casimir energy should be modified.
\end{abstract}
\newpage

\section {Introduction}

Recently, Verlinde \cite{verlinde00} has proposed that in the early
universe there exists a holographic bound on the sub-extensive entropy,
associated with the Casimir energy. When this bound is saturated, there emerges a formal coincidence between the Friedmann equation for $H^2 \equiv (\dot{a}/a)^2$ and the Cardy entropy formula \cite{cardy86, blote86}. The question that naturally arises, is whether this merging between the holographic principle, the entropy formula from conformal field theory, and the Friedmann equation from cosmology, is of deeper physical significance, and thus reproducable under more general conditions, or if it is just a formal coincidence. As one would expect, the Verlinde proposal has been the subject of study in cases where more general effects are accounted for; thus Wang et al. \cite{wang01} have considered universes having a cosmological constant, and Nojiri et al. \cite{nojiri01} have considered quantum bounds for the Cardy-Verlinde formula
 ( for other studies of related questions, see \cite{savonije01}).

The present paper focuses on one specific generalization of the original 
Verlinde setting, namely the presence of {\it viscosity} in the early 
universe. This case is clearly important as it corresponds to an attempt of 
 extending the formalism to non-conformally invariant theories. 
 The topic of viscous cosmology generally has attracted considerable interest recently. Moreover, we  allow in the latter part of the paper for the presence of a cosmological constant $\Lambda$, and we do not restrict the equation of state for the cosmic fluid to be necessary that of radiation dominance.

\section {Cardy-Verlinde formula in cosmology}
  Consider, then, the cosmic 
fluid whose four-velocity is 
$U^\mu=(U^0, U^i)$. In comoving coordinates, $U^0=1,~U^i=0$. In terms of the projection tensor $h_{\mu \nu}=g_{\mu \nu}+U_\mu U_\nu$ we can write the fluid's energy-momentum tensor as
\begin{equation}
T_{\mu\nu}=\rho U_\mu U_\nu+(p-\zeta \theta)h_{\mu\nu}-2\eta \sigma_{\mu\nu},
\end{equation}
\label{1}
assuming constant temperature in the fluid. Here, $\zeta$ is the bulk viscosity, $\eta$ the shear viscosity, $\theta \equiv {U^\mu}_{; \mu}$ the scalar expansion, and $\sigma_{\mu\nu}=h_\mu ^\alpha \,h_\nu ^\beta\, U_{(\alpha; \beta)}-\frac{1}{3}h_{\mu\nu} \,\theta$ the shear tensor  (more details are given, for instance, in \cite{brevik94, brevik00}). In accordance with common usage we omit henceforth the shear viscosity in view of the assumed complete isotropy of the fluid, although we have to mention that this is actually a nontrivial point. The reason is that the shear viscosity is usually so much greater than the bulk viscosity. Typically, after termination of the plasma era at the time of recombination ($T \simeq 4000$ K) the ratio $\eta / \zeta$ as calculated from kinetic theory is as large as about $10^{12}$ \cite{brevik94, brevik00}. Thus, even a slight anisotropy in the fluid would easily outweigh the effect of the minute bulk viscosity.

Assuming now a metric of the FRW type,
\begin{equation}
ds^2=-ds^2+a^2(t)\left( \frac{dr^2}{1-kr^2}+r^2 d\Omega^2 \right),
\end{equation}
\label{2}
with $k=-1,0,1$ the curvature parameter, we get $\theta =3\dot{a}/a \equiv 3H$. We define the effective pressure $\tilde{p}$ by
\begin{equation}
\tilde{p} \equiv p-\zeta \theta =p-3H\zeta.
\end{equation}
\label{3}
From Einstein's equations $R_{\mu\nu}-\frac{1}{2}Rg_{\mu\nu}+\Lambda g_{\mu\nu}=8\pi G\,T_{\mu\nu}$ we obtain the first Friedmann equation ("initial value equation")
\begin{equation}
H^2=\frac{8\pi G}{3}\,\rho+\frac{\Lambda}{3}-\frac{k}{a^2},
\end{equation}
\label{4}
where $\rho=E/V$ is the  energy density. This equation contains no viscous term. The second Friedmann equation ("dynamic equation"), when combined with Eq.~(4), yields
\begin{equation}
\dot{H}=-4\pi G (\rho +\tilde{p})+\frac{k}{a^2}.
\end{equation}
\label{5}
Here, the presence of viscosity is explicit.

We now recall that the entropy of a (1+1) dimensional CFT is given by the Cardy formula \cite{cardy86, blote86}
\begin{equation}
S=2\pi \sqrt{\frac{c}{6}\left(L_0-\frac{c}{24}\right)},
\end{equation}
\label{6}
where $c$ is the central charge and $L_0$ the lowest Virasoro generator.

  Let us assume that the universe is closed, and has a vanishing cosmological constant, $k=+1,~~~\Lambda=0$.
This is the case considered in  \cite{verlinde00} (in his formalism  the 
number $n$ of space dimensions is set equal to 3). The Friedmann equation (4) is seen to agree with the CFT equation (6) if we perform the substitutions
\begin{equation}
 L_0 \rightarrow \frac{1}{3}Ea,~~~c\rightarrow \frac{3}{\pi}\frac{V}{Ga},~~~S\rightarrow \frac{HV}{2G}.
\end{equation}
\label{7}
These substitutions are the same as in Ref.\cite{verlinde00}. We see thus that Verlinde's argument remains valid, even if the fluid possesses a bulk viscosity. Note that no assumptions have so far been made about the equation of state for the fluid.
Already a this point the following question thus naturally arises: is there a deeper connection between the laws of general relativity and those of quantum field theory?

Continuing this kind of reasoning, let us consider the three actual entropy definitions. First, there is the Bekenstein entropy \cite{bekenstein81},  $S_B=\frac{2\pi}{3}Ea$. The arguments for deriving this expression seem to be of a general nature; in accordance with Verlinde we find it likely that the Bekenstein bound $S \leq S_B$ is universal. We shall accept this expression for $S_B$ in the following, even when the fluid is viscous.

The next kind of entropy is the Bekenstein-Hawking expression $S_{BH}$, which is supposed to hold for systems with limited self-gravity: $S_{BH}=\frac{V}{2Ga}.$ Again, this expression relies upon the viscous-insensitive member (4) of Friedmann's equations. Namely, when $\Lambda=0$ this equation yields $S_B \lessgtr S_{BH}~~~~{\rm when} ~~~~Ha \lessgtr 1$. The borderline case between a weakly and a strongly gravitating system is 
thus at $Ha=1$. It is reasonable to identify $S_{BH}$ with the holographic 
entropy of a black hole with the size of the universe.

The third entropy concept is the Hubble entropy $S_H$. It can be introduced by starting from the conventional formula $A/4G$ for the entropy of a black hole. The horizon area $A$ is approximately $H^{-2}$, so that $S_H \sim H^{-2}/4G \sim HV/4G$ since $V\sim H^{-3}$. Arguments have been given by Easther and Lowe \cite{easther99}, Veneziano \cite{veneziano99}, Bak and Rey \cite{bak00}, and Kaloper and Linde \cite{kaloper99} for assuming the maximum entropy inside the universe to be produced by black holes of the size of the Hubble radius (cf. also \cite{fischler98}). According to Verlinde the FSB prescription (see \cite{verlinde00} for a closer discussion) one can determine the prefactor:  $S_H=\frac{HV}{2G}$. It is seen to agree with Eq.~(7).

One may now {\it choose} (see below) to define the Casimir energy $E_C$ as the violation of the Euler identity:
\begin{equation}
E_C \equiv 3(E+pV-TS)
\end{equation}
\label{8}
where, from scaling, the total energy $E$ can be decomposed as ($E_E$ is the extensive part) $E(S,V)=E_E(S,V)+\frac{1}{2}E_C(S,V)$. Due to  conformal invariance the products $E_E\, a$
and $E_C\, a$ are independent of the volume $V$, and a function of the
entropy $S$ only. From the known extensive behaviour of $E_E$ and the
sub-extensive behaviour of $E_C$ one may write (for CFT)
 \begin{equation}
E_E=\frac{C_1}{4\pi a} S^{4/3},~~~~E_C=\frac{C_2}{2\pi a} S^{2/3},
\end{equation}
\label{9}
where $C_1, C_2$ are constants whose product for CFTs is known: $\sqrt{C_1
C_2}=n=3$ (this follows from the AdS/CFT correspondence, cf. \cite{verlinde00}). From these expressions it follows that
\begin{equation}
S=\frac{2\pi a}{3}\sqrt{E_C(2E-E_C)}.
\end{equation}
\label{10}
This is the Cardy-Verlinde formula. Identifying $Ea$ with $L_0$ and $E_C\,a$ with $c/12$ we see that Eq.~(10) becomes the same as Eq.~(6), except from a numerical prefactor which is related to our assumption about $n=3$ space dimensions instead of $n=1$ as assumed in the Cardy formula.

The question is now: can the above line of arguments be carried over to
the case of a viscous fluid? The most delicate point here appears to be
the assumed pure entropy dependence of the product $Ea$. As we mentioned
above, this property was derived from conformal invariance, a property that is 
absent in the case under discussion. To examine
whether the property still holds when the fluid is viscous (and conformal
invariance is lost), we can start from the Friedmann equations (4) and (5), in the case $k=1, \, \Lambda=0$, and derive the "energy equation", which can be transformed to
\begin{equation}
\frac{d}{da}(\rho a^4)=(\rho -3\tilde{p})a^3.
\end{equation}
\label{11}
Thus, for a radiation dominated universe, $p=\rho /3$, it follows that
\begin{equation}
\frac{d}{dt}(\rho a^4)=\zeta \,\theta^2 a^4.
\end{equation}
\label{12}
Let us compare this expression, which is essentially the time derivative of the volume density of the quantity $Ea$ under discussion, with the four-divergence of the entropy current four-vector $S^\mu$. If $n$ is the number density and $\sigma$ the entropy per particle, we have $S^\mu=n\sigma U^\mu$ (we put $k_B=1$), which satisfies the relation (cf., for instance, Ref.~\cite{brevik94}) ${S^\mu}_{; \mu}=\frac{\zeta}{T}\theta^2$.
Since $(nU^\mu)_{;\mu}=0$ we have, in the comoving coordinate system, ${S^\mu}_{;\mu}=n\dot{\sigma}$, so that the time derivative of the entropy density becomes
\begin{equation}
n\dot{\sigma}=\frac{\zeta}{T}\theta^2.
\end{equation}
\label{13}
The two time derivatives (12) and (13) are seen to be proportional to $\zeta$. Since $\zeta$ is small, we can therefore insert for $a=a(t)$ the expression pertinent for a non-viscous, closed universe:$
a(t)=\sqrt{\frac{8\pi G}{3}\rho_0 a_0^4}\,\sin \eta$,
$\eta$ being the conformal time. Imagine now that Eqs.~(12) and (13) are integrated with respect to time. Then, since the densities $\zeta^{-1}\rho a^4$ and $\zeta^{-1} n\sigma$ can be drawn as functions of $t$, it follows that $\rho a^4$ can be considered as a function of $n\sigma$, or, equivalently, that $Ea$ can be considered as a function of $S$. We conclude that this property, previously derived on the basis of CFT, really appears to carry over to the viscous case.

The following point ought to be commented upon. The specific entropy $\sigma$ in Eq.~(13) is the usual thermodynamic entropy per particle. The identification of $S$ with $HV/(2G)$, as made in Eq.~(7), is however something different, since it is derived from  a comparison with the Cardy formula (6).  Since this entropy is the same as the Hubble entropy $S_H$ we can write the equation as 
$ n\sigma_H=H/(2G), $
where  $\sigma_H$ is the Hubble entropy per particle. This quantity is different from  $\sigma$, since it does not follow from thermodynamics plus Friedmann equations alone, but from the holographic principle. The situation is actually not peculiar to viscous cosmology. It occurs if $\zeta=0$ also. The latter case is easy to analyze analytically, if we focus attention on the case $t\rightarrow 0$.  Then, for any value of $k$, we have $a \propto t^{1/2}$, implying  that $H=1/(2t)$. Moreover, from the equation of continuity,  $(nU^\mu)_{;\mu}=0$,  which for a FRW universe yields  $na^3=constant$, so that $n \propto t^{-3/2}$.  The above equation for $\sigma_H$  then yields
    $\sigma_H \propto t^{1/2}$. This is obviously different from the result for the thermodynamic entropy $\sigma$: from Eq.~(13) we simply get $\sigma=constant$ when $\zeta=0$. The two specific entropies are thus different even in this case.

\section {Casimir energy}
   Let us make a couple of general remarks on the 
above formalism.
They are based on physical, rather than mathematical, considerations, and 
are not primarily concerned with viscosity.
First, one may wonder about the legitimacy of defining the Casimir energy such as in Eq.~(8). Usually, within the Green function approach, in a spherical geometry the Casimir energy is calculated indirectly, by integration of the Casimir surface force density $f=-(1/4\pi a^2)\partial E/\partial a$. The force $f$ in turn is calculated by first subtracting off the volume-dependent parts of the two scalar Green functions; this is in agreement with the physical requirement that $f \rightarrow 0$ at $r \rightarrow \infty$. (The typical example of this configuration is that of a conducting shell; cf., for instance, Ref.~\cite{milton78}.)
 That this kind of procedure should lead to the same result as Eq.~(8), 
which merely expresses a violation of the thermodynamic Euler identity, 
is in our opinion not evident.

Our second remark is about the physical meaning of taking the Casimir energy $E_C$ to be {\it positive}. Verlinde assumes that $E_C$ is bounded by the total energy $E$: $
E_C \leq E.$ This may be a realistic bound for some of the CFTs. However, in general cases, 
it is not true.
For a realistic dielectric material it is known that the full Casimir energy is not positive; the dominant terms in $E_C$ are definitely {\it negative}. From a statistical mechanical point of view this follows immediately from the fact that the Casimir force is the integrated effect of the attractive van der Waals force between the molecules. Now the case of a singular conducting shell is  complicated - there are two limits involved, namely the infinitesimal thickness of the shell and also the infinite conductivity (or infinite permittivity) - and a microscopical treatment of such a configuration has to our knowledge not been given. What {\it is} known, is the microscopical theory for a dielectric ball. Let us write down, for illustration, the expression derived by Barton \cite{barton99} for a dilute ball:
\[ E_C=-\frac{3\gamma}{2\pi^2}\frac{V}{\lambda^4} \]
\begin{equation}
+\gamma^2 \left( -\frac{3}{128 \pi^2}\frac{V}{\lambda^4}+\frac{7}{360\pi^3}\frac{A}{\lambda^3}
-\frac{1}{20\pi^2}\frac{1}{\lambda}+\frac{23}{1536\pi}\frac{1}{a} \right),
\end{equation}
\label{14}
where $\gamma =(\epsilon-1)/\epsilon,~~A$ is the surface area, and $\lambda$ is a cutoff parameter. This expression, derived from quantum mechanical perturbation theory, agrees with the statistical mechanical calculation in Ref.~\cite{hoye00}, and also essentially with Ref.~\cite{bordag99} (there are some numerical factors different in the cutoff-dependent terms). It is evident from this expression that the dominant, cutoff, dependent voume terms, are negative.

We see that there remains one single, cutoff independent, term in Eq.~(14). This term is in fact positive. It can be derived  from macroscopic electrodynamics also,  by using either dimensional continuation or zeta-function regularization, as has been done in Ref.~\cite{brevik99}. In the present context the following question becomes however natural: how can a positive, small, cutoff dependent term in the Casimir energy play a major role in cosmology?
In another words, why should the matter necessarily be conformal? Of
course, our universe is different from a dielectric ball, and we are
not simply stating that Verlinde's method is incorrect.
 Our aim is merely to stress the need of caution, when  results from one field in physics are applied to another field.
In any case, all this suggests that the consideration of non-conformally 
invariant situations should significally change the dynamical entropy 
bounds and bounds for Casimir energy.

\section {Discussion} 
 We round off this paper by making three brief remarks.

({\it i})  Our treatment above, in Sec.~2, was based upon the set of cosmological assumptions $ \{ p=\rho/3,\, k=+1, \,\Lambda=0 \}.$ The recent development of Wang et al. \cite{wang01} is interesting, since it allows for a nonvanishing cosmological constant (still assuming a closed model). One of the scenarios treated in \cite{wang01} is that of a de Sitter universe ($\Lambda >0$) occupied by a universe-sized black hole. A black hole in de Sitter space has the metric
$ ds^2=-f(r)dt^2+f^{-1}(r)dr^2+r^2 d\Omega^2,$
where $f(r)=1-2MG/r-\Lambda r^2/3$. The region of physical interest is that lying between the inner black hole horizon and the outer cosmological horizon, the latter being determined by the magnitude of $\Lambda$. 

Although we do not enter into any detail about this theory, we make the following observations: the above metric is {\it static}; there is no time-dependent scale factor involved, and the influence from viscosity will not turn up in the line element. Moreover, Wang et al. make use of only the member (4) of Friedmann's equations which, as we have noticed, is formally independent of viscosity.

Does this imply that viscosity is without any importance for the present kind of theory? The answer in our opininon is no, since the theory operates implicitly with the concept of the maximum scale factor $a_{max}$ in the closed Friedmann universe. In order to calculate $a_{max}$, one has to solve the Friedmann equation (5) also, which contains the viscosity through the modified pressure $\tilde{p}$. Thus, viscosity comes into play after all, though in an indirect way.

({\it ii})  We note that there exists a physically interesting variant of the theory of black holes, namely the one proposed by 't Hooft \cite{hooft98}. This model treats the black hole as a system endowed with an envelope of matter obeying a given equation of state, the matter being treated as a {\it source} in Einstein's equations. In our context the most interesting aspect of the 't Hooft model is the calculation of the entropy of the self-screening envelope of matter: in order to obtain the Hawking formula $S=A/4G$ one has to choose the Zel'dovich state equation $p=\rho$, corresponding to the velocity of sound being equal to the velocity of light.

Thus, whereas in the preceding we calculated the Hubble entropy $S_H$ starting from the conventional Hawking entropy formula, maintaining the usual assumption about a radiation dominated universe satisfying $p=\rho/3$, the same result for the entropy follows from the 't Hooft black hole model if we take the cosmic fluid to be a Zel'dovich fluid. The Zel'dovich equation of state may appear rather peculiar, but it is natural to recall here that analogous "equations of state" are found to lead to significant simplifications of the formalism in other areas of physics also, notably in the Casimir theory in spherical media. Thus, if the medium satifies the condition $\epsilon \mu =1$ where $\epsilon$ is the permittivity and $\mu$ the permeability, then the difficulties that one otherwise has in constructing the contact term in the Green functions go away \cite{brevik82}. 

({\it iii})  One may ask:  how can the definition of Casimir energy (and entropy bounds)
 be modified for
non-conformal matter? This is clearly important as it may
connect the Casimir effect theory with non-conformal extension of the AdS/CFT
correspondence. Moreover, it is not expected that any symmetry at the
early universe is exact, so it is hard to imagine that there are no
violations of conformal invariance.
Second, these results may be easily generalized to account for quantum
effects.
In particular, using the results of  Ref.~\cite{nojiri01}, one sees that quantum effects only induce a
non-trivial effective cosmological constant. This effective cosmological
constant may play a major role in  viscous cosmology.

\end{document}